\documentclass[letterpaper, 10 pt, journal, twoside]{ieeetran}                                     

\usepackage{amsmath, bm}
\usepackage{amssymb}
\usepackage{commath}
\usepackage{graphicx}

\newtheorem{theorem}{Theorem}[section]                               

\newtheorem{definition}{Definition}[section]
\newtheorem{remark}{Remark}
\newtheorem{corollary}{Corollary}[section]

\newcounter{as}
\newenvironment{assum}
{\par\noindent
	\refstepcounter{as}%
	A\theas.}~\itshape\ignorespaces
{\par\ignorespacesafterend}

\begin{document}

\title{Consistent Estimators for Nonlinear Vessel Models} 

\author{Fredrik Ljungberg and Martin Enqvist
	\thanks{This work was supported by the Vinnova Competence Center LINK-SIC. The authors are with the Department of Electrical Engineering,
		Linköping University, 58183 Linköping Sweden (email: \tt\footnotesize fredrik.ljungberg@liu.se; \tt\footnotesize martin.enqvist@liu.se).}
}

\maketitle
\thispagestyle{empty}
\pagestyle{empty}

\begin{abstract}                          
	In this work, the issue of obtaining consistent parameter estimators for nonlinear regression models where the regressors are second-order modulus functions is explored.  It is shown that consistent instrumental variable estimators can be obtained by estimating first and second-order moments of non-additive environmental disturbances' probability distributions as nuisance parameters in parallel to the sought-after model parameters, conducting experiments with a static excitation offset of sufficient amplitude and forcing the instruments to have zero mean. The proposed method is evaluated in a simulation example with a model of a marine surface vessel.
\end{abstract}

\section{Introduction}
Greybox identification is the practice of identifying dynamical systems based on data while exploiting partial prior information, see for example \cite{bohlin2006practical} for a comprehensive treatment of greybox identification for industrial processes. Another application where greybox identification is commonly employed is robotics. Regarding identification of vessels, $i.e.$, mobile robots, the prior information is usually given in terms of physical knowledge and is used to form valid and understandable model structures. The time-invariant parameters of these model structures, which are not already known, are then adapted to collected data.

The equations describing the dynamics of robots in motion are often based on Newton's second law, where the force equations are given by
\begin{align} \label{eq:newton2}
	\bm{F} &= m \dot{\bm{V}} + \bm{\omega} \times m \bm{V}.
\end{align}
Here $\bm{F}$ is the sum of forces affecting the robot, caused by $e.g.$, propulsion, drag and environmental disturbances such as wind, currents and gravity. Moreover, $m \dot{\bm{V}}$ comes from momentum change and the term $\bm{\omega} \times m \bm{V}$ constitutes Coriolis and centripetal effects, which need to be accounted for since the motion is described in a robot-fixed coordinate system that is itself in movement relative to an inertial frame.

Basic motions that do not deviate too much from an operating point are usually well-described by linearized relationships.  However, to accurately describe motion in wider operating regions, it is sometimes necessary to consider nonlinear models. Nonlinear behavior appears in many engineering problems, not least in robotics where, for example, nonlinear drag and Coriolis effects make linear models imprecise and unable to reproduce essential aspects of system behaviors. There are many ways to describe the drag or resistance of an object in a fluid environment such as air or water, see for example \cite{long1999velocity}. A fairly simple description that often works well is to assume that the drag force is proportional to the squared velocity. The Coriolis and centripetal effects are usually well-described by quadratic functions as well. This means that more general system descriptions can be obtained if, in addition to linear ones, quadratic regressors are allowed in the model structures.

Besides actuators, unknown external forces affect the steering dynamics. Dealing with these typically quite impactful process disturbances correctly during model estimation is quite challenging already in the linear case and becomes even more difficult when models are nonlinear. If the measurement data is collected under the presence of process disturbances and these are not accounted for during the model estimation, the resulting model might be biased. In practical terms, this means that instead of just describing the sought characteristics of the vessel in question, the model can adapt to the conditions prevailing under the data acquisition. Moreover, there is always uncertainty associated with measuring something. Dealing with this inherent uncertainty is also of importance to obtain accurate models.

Consequently, in addition to increased possibilities of obtaining accurate system descriptions, the use of nonlinear models comes at a price. In fact, accurate parameter estimation for general nonlinear model structures, using data collected under the presence of generic disturbances, is a fairly open problem, see for example~\cite{ljung2010perspectives}. As a consequence, there is a substantial effort focused on the problem. In~\cite{wernholt2011nonlinear}, a two-step solution was proposed where frequency-response estimates for local linear models were first estimated for a chosen number of operating points. Suitable values for the parameters of a nonlinear model were then found in a second step by minimizing the discrepancy between these locally valid frequency-response estimates and the parametric frequency response of the nonlinear model. Another common way of approaching the issue of parameter estimation for nonlinear models is to consider cases where the Maximum Likelihood problem can be formulated and solved. This was done using the Expectation-Maximization algorithm and particle smoothing in~\cite{schon2011system}. Another approach is a prediction-error method~(PEM). In~\cite{abdalmoaty2019identification}, a prediction-error perspective with suboptimal predictors was explored. The results showed that linear predictors can give consistent estimators in a prediction-error framework for a quite large class of nonlinear models, even when data are generated from systems with non-additive disturbances. Moreover, the benefits of supplementing PEM with a parameterized linear observer for capturing unmodelled disturbance characteristics were investigated in~\cite{larsson2019system}. This linear observer turned out to be a quite easily accessible way of compensating for miss-specified predictors. 

In general, formulating the ML problem for parameter estimation or constructing a predictor to be used in PEM, requires prior knowledge about the disturbances' probability distributions. If the disturbances are not well-described as white Gaussian processes, the required prior knowledge is not even necessarily restricted to first and second-order moments. Environmental disturbances such as wind and water currents, which typically affect robots in motion, can show varying behavior during a data collection experiment, for example, due to turbulent flows. Therefore, to not have to make hard assumptions regarding the disturbances' character, the parameter estimators developed in this work are based on the instrumental variable (IV) method. Moreover, all the aforementioned works deal with quite general model classes while in this work, the focus is on developing consistent parameter estimators for a special class of nonlinear models called second-order modulus models. These models do, as the name suggests, include second-order terms. Additionally, the model class allows the use of the modulus function. The motivation for this is that in some applications models should be based on odd functions for symmetry purposes, which for quadratic nonlinearities can be resolved by using absolute values. This model class was first proposed for ships in \cite{fedyaevsky1964control} but includes models used in a variety of other robotic applications, $e.g.$, underwater vehicles \cite{karras2018unsupervised}, hovercrafts \cite{xie2018robust}, fixed-wing aircraft \cite{salman2006attitude}, multirotor aircraft \cite{zhang2014survey} and blimps \cite{zufferey2006flying}. Vessels of all these types are affected by forces and moments according to Newton’s laws and their motions are bound by kinematic relationships. Typically, both the dynamic and the kinematic equations are nonlinear but a key difference is that the unknown physical properties, and consequently the parameters that need to be estimated, are present in the dynamic relations, which are well-described by second-order modulus models. Effects such as wind, currents and gravity can then be treated as process disturbances.

This work complements earlier works in \cite{ljungberg2019,ljungberg2020} by acknowledging kinematic dependencies between non-additive disturbance effects and system states, which make the disturbances harder to deal with using an IV method alone. The proposed solution for this is to estimate the first and second-order moments of the non-additive environmental disturbances' probability distributions as nuisance parameters in parallel to the sought-after model parameters. Additionally, the extension to second-order modulus systems with two non-additive disturbances is addressed. These extensions make the proposed estimation framework more useful in practice and its potential is shown with a realistic simulation example.

\section{Problem formulation and preliminaries}
For describing a general second-order modulus model, it is convenient to first define a second-order modulus function. 

\begin{definition} \label{def:modulus} A second-order modulus function is a function, $\bm{f}_{\text{som}} \ : \ \mathbb{R}^{n_x+ n_{\theta}} \rightarrow \mathbb{R}^{n_f} $ that can be written as 
\begin{align*}
	\bm{f}_{\text{som}}(\bm{x}, \bm{\theta}) = \bm{\Phi}^T(\bm{x})  \bm{\theta},
\end{align*}
where each element of the $n_{\theta} \times n_f$ matrix $\bm{\Phi}(\bm{x})$ is on one of the forms $x_i$, $\abs{x_i}$, $x_i x_j$, $x_i \abs{x_j}$ for $i, j \leq n_x$ or zero and $\bm{\theta} \in \mathbb{R}^{n_{\theta}}$ is a vector of coefficients. 
\end{definition}

Now, consider the system
\begin{subequations} \label{eq:system}
\begin{align}
	\bm{x}(k+1) &= \bm{f}_{\emph{som}}\left( \begin{bmatrix}  \bm{x}(k) + \bm{R}(k)\bm{v}(k) \\ \bm{u}(k)  \end{bmatrix}, \bm{\theta}_0 \right) + \bm{w}(k), \\
	\bm{y}(k) &= \bm{x}(k) + \bm{e}(k), \\
	\bm{Y}_R(k) &= \bm{R}(k) + \bm{E}_R(k),
\end{align}
\end{subequations}
where the vector $\bm{u}(k)\in \mathbb{R}^{n_u}$ is a known input signal and $\bm{x}(k) \in \mathbb{R}^{n_x}$ is a vector consisting of the latent system states, all of which are measured directly (with noise) and collected in the output vector $\textbf{y}(k) \in \mathbb{R}^{n_x}$. Moreover, $\bm{R}(k) \in \mathbb{R}^{n_x \times n_v}$ is a time-varying matrix, $\textbf{v}(k) \in \mathbb{R}^{n_v}$ and $\textbf{w}(k)  \in \mathbb{R}^{n_x}$ are external signals that are assumed to be unknown (process disturbances) while $\textbf{e}(k) \in \mathbb{R}^{n_x}$ and $\bm{E}_R(k) \in \mathbb{R}^{n_x \times n_v}$ constitute additive measurement errors, which are also assumed to be unknown. The system is described by the parameter vector $\bm{\theta}_0 \in \mathbb{R}^{n_{\theta}}$, which does not vary over time. In addition to this, the following premises regarding the system are assumed to be imposed. 

\begin{assum} \label{ass:second_order_modulus_system}
	$\bm{f}_{\emph{som}}(\cdot)$ is a second-order modulus function in agreement with Definition~\ref{def:modulus} and its structure is known.
\end{assum} 

\begin{assum} \label{ass:e_noise}
	The measurement noises, $\bm{e}(k)$ and $\bm{E}_R(k)$, are stationary signals with zero mean and well-defined moments of any order.
\end{assum}

\begin{assum} \label{ass:w_noise}
	The process disturbance $\bm{w}(k)$ is a stationary signal with well-defined moments of any order.
\end{assum}

\begin{assum} \label{ass:v_noise} 
	The process disturbance $\bm{v}(k)$ is a stationary signal with well-defined moments of any order. 
\end{assum}

%

\begin{assum} \label{ass:eta_bounds}
	 The signals $\bm{e}(k)$, $\bm{E}_R(k)$, $\bm{R}(k)$ and $\bm{v}(k)$ are bounded in magnitude such that
	 \begin{align*}
	 	-\bm{\eta}_v \leq \bm{R}(k) \bm{v}(k) \leq \bm{\eta}_v, \\
	 	-\bm{\eta}_e \leq \bm{e}(k) + \left(\bm{R}(k) + \bm{E}_R(k)\right) E\left\{ \bm{v}(k) \right\} \leq \bm{\eta}_e.
	 \end{align*}
\end{assum}

Moreover, given that $N_E$ experiments are performed, where in each $N_D$ data points are collected, the following assumptions are made regarding the experiment design.

\begin{assum} \label{ass:open_loop}
	The system is operating in open loop, $i.e.$, the input, $\bm{u}(k)$, does not depend on the measurements, $\bm{y}(k)$ and $\bm{Y}_R(k)$, and is consequently assumed to be independent of all disturbances. This in turn means that $\bm{x}(k)$ and $\bm{R}(k)$ are assumed to be independent of the measurement noises, $\bm{e}(k)$ and $\bm{E}_R(k)$.
\end{assum}

\begin{assum} \label{ass:excitation_offset}
	The input in each experiment is such that it excites the system to the extent that each of its states, $x_1(k), \ \hdots \ x_{n_x}(k)$, continuously has an amplitude that is sufficiently well-separated from the origin
	\begin{align*}
		\abs{x_i(k)} > \text{max}(\eta_{e, i}, \eta_{v, i}),
	\end{align*}
	for $k = 1, \ \hdots \ N_D$, $i = 1, \ \hdots \ n_x$.
\end{assum}

Performing experiments by A\ref{ass:excitation_offset} was a key step proposed in \cite{ljungberg2019} and is central for the ideas presented here as well. The assumption makes it possible to temporarily treat second-order modulus functions as normal second-order functions during the analysis of the parameter estimation.

\subsection{The IV estimator}
The suggested estimators will be based on the IV method. For a predictor model, $\hat{\bm{y}}(k, \bm{\theta})$, parameterized in $\bm{\theta}$, the IV estimate is defined as
\begin{align} \label{eq:IV_estimator}
	\hat{\bm{\theta}}_N^{IV} = \text{sol} \left \{ \frac{1}{N}\sum_{k=1}^N \bm{Z}(k) \left(\bm{y}(k)-\hat{\bm{y}}(k, \bm{\theta})\right) = 0 \right\},
\end{align}
where $\textbf{Z}(k)$ is called the instrument matrix and the notation $\text{sol} \left \{ \bm{f}(\textbf{x}) = 0 \right\}$ is used for the solution to the system of equations \mbox{$f_i(\textbf{x}) = 0, \ i = 1, \hdots, \ n_x$}. Provided that the model can be written on regression form, $\hat{\bm{y}}(k, \bm{\theta}) = \bm{\Phi}^T(k) \bm{\theta}$, the IV estimator will be consistent if
\begin{subequations}
	\begin{align}
		\bar{E}\left\{ \bm{Z}(k) \bm{\Phi}^T(k)\right\} \ \text{is full rank}, \label{eq:IV_cond1} \\
		\bar{E}\left\{ \bm{Z}(k) \left( \bm{y}(k) - \hat{\bm{y}}(k, \bm{\theta}_0) \right)\right\} = 0. \label{eq:IV_cond2}
	\end{align}
\end{subequations}
Here the notation $\Bar{E}\{\cdot\} = \lim_{N \to \infty} \frac{1}{N} \sum_{k=1}^N E \{\cdot \}$ was adopted from \cite{ljung1999system}. See \cite{soderstrom1988system} for more details regarding IV methods.

\subsection{Basic predictor}
Following Definition~\ref{def:modulus}, the system dynamics of \eqref{eq:system} can be expressed as
\begin{align*} 
	\bm{x}(k+1) &= \bm{\Phi}^T\left( \begin{bmatrix}  \bm{x}(k) + \bm{R}(k)\bm{v}(k) \\ \bm{u}(k)  \end{bmatrix} \right)  \bm{\theta}_0 + \bm{w}(k),
\end{align*}
and since the structure of this system is known by A\ref{ass:second_order_modulus_system}, a simple way of modelling it is to consider the predictor
\begin{align} \label{eq:basic_predictor}
	\hat{\bm{y}}_1(k, \bm{\theta}) &= \bm{\Phi}^T\left( \begin{bmatrix} \bm{y}(k-1) \\ \bm{u}(k-1)  \end{bmatrix} \right)  \bm{\theta}.
\end{align}
The possibilities of obtaining consistent parameter estimators for $\bm{\theta}$ with this basic predictor using IV methods have been explored in earlier works, \cite{ljungberg2019,ljungberg2020}. Due to the errors-in-variables formulation in \eqref{eq:system}, $i.e.$, the fact that the state is measured with uncertainty and not known exactly, $\bm{e}(k) \neq 0$, as well as the non-additive process disturbance, $\bm{v}(k) \neq 0$, the conditions \eqref{eq:IV_cond1} and \eqref{eq:IV_cond2} are challenging to fulfill at the same time. One way of doing so is to consider an experiment with excitation offset and an instrument matrix with zero mean. This was in \cite{ljungberg2020} shown to give consistent estimators when $E\{\bm{v}(k)\} = 0$ but not otherwise.

\subsection{Augmented predictor}
In this work, the usefulness of an augmented set of predictors that are better suited to deal with $E\{\bm{v}(k)\} \neq 0$ will be explored. By A\ref{ass:second_order_modulus_system}, the matrix
\begin{align*}
\bm{\Phi}^T\left( \begin{bmatrix}  x_1(k) + \sum_{j=1}^{n_v} \bm{R}_{1,j}(k) v_j(k) \\ \vdots \\ x_{n_x}(k) + \sum_{j=1}^{n_v} \bm{R}_{n_x,j}(k) v_j(k) \\ u_1(k) \\ \vdots \\ u_{n_u}(k) \end{bmatrix} \right), 
\end{align*}
will consist of a combination of different kinds of second-order modulus elements. Moreover, by A\ref{ass:eta_bounds} and A\ref{ass:excitation_offset}, in each experiment, $i  = 1, \hdots, N_E$, either of the conditions
\begin{align*}
	x_{\ell}(k) + \sum_{j=1}^{n_v} [\bm{R}(k)]_{\ell,j}  v_j(k)  > 0, \ \forall \ k = 1, \hdots, N_D, \\
	x_{\ell}(k) + \sum_{j=1}^{n_v} [\bm{R}(k)]_{\ell,j}  v_j(k)  < 0, \ \forall \ k = 1, \hdots, N_D,
\end{align*}
will hold for $\ell = 1, \hdots, n_x$. This, together with the fact that the input is perfectly known, means that all absolute values can be removed and consequently that $\bm{\Phi}(\cdot)$ can be treated as a regular (matrix-valued) second-order function during the parameter estimation. When the squares of this function are developed, a limited number of terms can appear. Some will not depend on $\bm{v}(k)$ whereas some will include either a factor  $v_j(k)$ or a factor $v_j(k) v_m(k)$ for some $j,m = 1, \hdots, n_v$. Furthermore, once $\bm{\Phi}^T(\cdot)$ is multiplied with $\bm{\theta}_0$, each term will include a factor $\theta_{0,p}$ for some $p = 1,\hdots, n_{\theta}$. Collecting these factors in vectors makes it possible to express the system dynamics as
\begin{align*}
	\textbf{x}(k+1) &= \bm{\Phi}^T\left(\bm{\Omega}_1(k)\right)  \bm{\theta}_0 + \bm{\Phi}_{\rho, i}^T\left(\bm{\Omega}_2(k)\right) \underbrace{\begin{bmatrix}  \vdots \\ v_j(k) \theta_{0,p} \\ \vdots \end{bmatrix}}_{= \bm{\Omega}_{\rho}(k)} + \\
	& \bm{\Phi}_{\lambda, i}^T\left(\bm{\Omega}_3(k)\right) \underbrace{\begin{bmatrix}  \vdots \\ v_j(k) v_{m}(k) \theta_{0,p} \\ \vdots \end{bmatrix}}_{= \bm{\Omega}_{\lambda}(k)} + \textbf{w}(k),
\end{align*}
for each experiment $i$. Here $\bm{\Phi}_{\rho, i}(\cdot)$ and $ \bm{\Phi}_{\lambda, i}(\cdot)$ are element-wise quadratic functions and
\begin{align*}
	\bm{\Omega}_1(k) \! = \! \begin{bmatrix}  \textbf{x}(k) \\ \textbf{u}(k)  \end{bmatrix}\!\!, \ \bm{\Omega}_2(k) \! = \! \begin{bmatrix}  \textbf{x}(k) \\ \textbf{u}(k) \\ \! \bm{\Omega}_3(k) \! \end{bmatrix}\!\!, \
	\bm{\Omega}_3(k) \!= \! \text{vec}\left( \bm{R}(k) \right).
\end{align*}
Now, consider the predictors
\begin{align} \label{eq:augmented_predictor}
	\hat{\bm{y}}_{2,i}(k, \bm{\theta},  \bm{\rho}, \bm{\lambda}) &= \begin{bmatrix} \bm{\Phi}\left(\bm{\Omega}'_1(k-1)\right) \\ \bm{\Phi}_{\rho, i}\left(\bm{\Omega}'_2(k-1)\right)  \\ \bm{\Phi}_{\lambda, i}\left(\bm{\Omega}'_3(k-1)\right) \end{bmatrix}^T \begin{bmatrix} \bm{\theta} \\  \bm{\rho} \\  \bm{\lambda}  \end{bmatrix},
\end{align}
for $i  = 1, \hdots, N_E$, where
\begin{align*}
	\bm{\Omega}'_1(k) \! = \! \begin{bmatrix}  \textbf{y}(k) \\ \textbf{u}(k)  \end{bmatrix}\!\!, \ \bm{\Omega}'_2(k) \! = \! \begin{bmatrix}  \textbf{y}(k) \\ \textbf{u}(k) \\ \! \bm{\Omega}'_3(k) \! \end{bmatrix}\!\!, \
	\bm{\Omega}'_3(k) \!= \! \text{vec}\left( \bm{Y}_R(k) \right),
\end{align*}
and where, in addition to the model parameters $\bm{\theta}$, the vectors~$\bm{\rho}$ and $\bm{\lambda}$ are left as free variables during estimation. In this work, the consistency of an IV estimator for the predictors defined by \eqref{eq:augmented_predictor} will be analyzed and it will be shown that~$\bm{\rho}$ will converge to a function of the first-order moment of the non-additive disturbance,~$E\{ \bm{v}(k) \} = \bar{\bm{v}}$, as the number of data points increases, whereas $\bm{\lambda}$ asymptotically will depend on the second-order moment of~$\bm{v}(k)$. Regarding the model structure, the following assumption is made. 

\begin{assum} \label{ass:global_identifiability}
	The vectors $\bm{\theta}$, $\bm{\rho}$ and $\bm{\lambda}$ are globally identifiable in \eqref{eq:augmented_predictor} according to the definition in~\cite{ljung1999system}.
\end{assum}

Note that $\bm{\Phi}_{\bm{\rho}, i}(k) \in \mathbb{R}^{n_{\rho} \times n_x}$,  $\bm{\Phi}_{\bm{\lambda}, i}(k) \in \mathbb{R}^{n_{\lambda} \times n_x}$ and consider the case where, for each experiment,~$i$, there is an instrument matrix, $\bm{Z}_i(k) \in \mathbb{R}^{(n_{\theta} + n_{\rho} + n_{\lambda}) \times n_x}$, that fulfills the following assumptions.
\begin{assum} \label{ass:z_indep} 
	$\textbf{Z}_i(k)$ is independent of the disturbance signals $\bm{e}(k)$, $\bm{E}_R(k)$ and $\bm{w}(k)$.
\end{assum}

\begin{assum} \label{ass:Ez=0}
	$\Bar{E}\{ \textbf{Z}_i(k) \} = 0$ and all moments of higher order are well-defined.
\end{assum}

\begin{assum} \label{ass:informative_data} The matrix
\begin{align*}
\bm{H} &= \begin{bmatrix} \bm{H}_1 & \hdots & \bm{H}_{N_E} \end{bmatrix}^T \in \mathbb{R}^{N_E(n_{\theta} + n_{\rho} + n_{\lambda}) \times (n_{\theta} + n_{\rho} + n_{\lambda})},
\end{align*}
with block elements
	\begin{align*}
		\bm{H}_i &= \Bar{E} \left\{ \bm{Z}_i(k) \begin{bmatrix} \bm{\Phi}\left(\bm{\Omega}'_1(k-1)\right) \\ \bm{\Phi}_{\rho, i}\left(\bm{\Omega}'_2(k-1)\right)  \\ \bm{\Phi}_{\lambda, i}\left(\bm{\Omega}'_3(k-1)\right) \end{bmatrix}^T \right\},
	\end{align*}
has full rank.
\end{assum}

The final assumption declares that when $N_D \to \infty$, the parameters can be determined uniquely, $i.e.$, that the data from all the experiments combined are sufficiently informative. It can be noted that A\ref{ass:global_identifiability} is a necessary condition for A\ref{ass:informative_data} to hold.

Since an exact solution to \eqref{eq:IV_estimator} might not exist, the IV estimate is obtained as the least-squares solution to the system of $(n_{\theta} + n_{\rho} + n_{\lambda})N_E$ equations
\begin{align} \label{IV_problem}
	\begin{cases} \frac{1}{N_D}\sum\limits_{k=1}^{N_D} \bm{Z}_1(k) \left(\bm{y}(k)-\hat{\bm{y}}_{2,1}(k, \bm{\theta}, \bm{\rho}, \bm{\lambda})\right) = 0, \\ \qquad \qquad \qquad \qquad \qquad\vdots \\  \frac{1}{N_D} \hspace{-0.4cm} \sum\limits_{k=(N_E-1)N_D+1}^{N_E N_D} \hspace{-0.6cm} \bm{Z}_{N_E}(k) \left(\bm{y}(k)-\hat{\bm{y}}_{2, N_E}(k, \bm{\theta}, \bm{\rho}, \bm{\lambda})\right) = 0 .  \end{cases}
\end{align}

\begin{theorem} \label{theorem1}
	Provided that A\ref{ass:second_order_modulus_system}-A\ref{ass:informative_data} are fulfilled, the IV method defined by \eqref{IV_problem} is a consistent estimator of $\bm{\theta}_0$.
\end{theorem}

\begin{IEEEproof}
An IV estimator is consistent if conditions \eqref{eq:IV_cond1} and \eqref{eq:IV_cond2} hold. By A\ref{ass:informative_data}, it is immediately possible to conclude that \eqref{eq:IV_cond1} is fulfilled. Consequently, it remains to show that \eqref{eq:IV_cond2} holds. 

For each experiment, $i  = 1, \hdots, N_E$, the model residual, $\bm{y}(k) - \hat{\bm{y}}_{2,i}(k, \bm{\theta}, \bm{\rho}, \bm{\lambda})$, can by A\ref{ass:second_order_modulus_system}, A\ref{ass:eta_bounds} and A\ref{ass:excitation_offset} be expressed~as
\begin{align*}
	\begin{bmatrix} \bm{\Phi}\left(\bm{\Omega}_1(k-1)\right) \\ \bm{\Phi}_{\rho, i}\left(\bm{\Omega}_2(k-1)\right)  \\ \bm{\Phi}_{\lambda, i}\left(\bm{\Omega}_3(k-1)\right)  \end{bmatrix}^T & \begin{bmatrix} \bm{\theta}_0 \\  \bm{\Omega}_{\rho}(k-1) \\  \bm{\Omega}_{\lambda}(k-1)  \end{bmatrix} + \bm{w}(k-1) \\
	&\hspace{-5mm}+ \bm{e}(k) - \begin{bmatrix} \bm{\Phi}\left(\bm{\Omega}'_1(k-1)\right) \\ \bm{\Phi}_{\rho, i}\left(\bm{\Omega}'_2(k-1)\right)  \\ \bm{\Phi}_{\lambda, i}\left(\bm{\Omega}'_3(k-1)\right) \end{bmatrix}^T \begin{bmatrix} \bm{\theta} \\  \bm{\rho} \\  \bm{\lambda}  \end{bmatrix},
\end{align*}
where the elements of the regression matrices are quadratic functions. When the squares are developed, it is possible to express $\bm{\Phi}\left(\bm{\Omega}'_1(k)\right)$ as
\begin{align*}
	\bm{\Phi}\left(\bm{\Omega}_1(k)\right) + \bm{f}_1\left(\bm{\Omega}_1(k), \bm{e}(k) \right) + \bm{f}_2\left(\bm{e}(k)\right),
\end{align*}
where each element of $\bm{f}_1(\cdot,\cdot)$ is bilinear in the arguments and $\bm{f}_2(\cdot)$ is an element-wise quadratic function. Similarly, $\bm{\Phi}_{\rho, i}\left(\bm{\Omega}'_2(k)\right)$ can be expressed as
\begin{align*}
	\bm{\Phi}_{\rho, i}\left(\bm{\Omega}_2(k)\right) + \bm{f}_3\left(\bm{\Omega}_2(k), \bm{e}'(k)\right) + \bm{f}_4\left(\bm{e}'(k)\right),
\end{align*}
where each element of $\bm{f}_3(\cdot,\cdot)$ is bilinear in the arguments, $\bm{f}_4(\cdot)$ is an element-wise quadratic function and $\bm{e}'(k) = \begin{bmatrix}  \bm{e}^T(k) & \text{vec}\left( \bm{E}_R(k) \right)^T \end{bmatrix}^T$. Lastly, $\bm{\Phi}_{\lambda, i}\left(\bm{\Omega}'_3(k)\right)$ can be expressed as
\begin{align*}
	\bm{\Phi}_{\lambda, i}\left(\bm{\Omega}_3(k)\right) + \bm{f}_5\left(\bm{\Omega}_3(k), \bm{e}''(k)\right)  + \bm{f}_6\left(\bm{e}''(k) \right),
\end{align*}
where each element of $\bm{f}_5(\cdot,\cdot)$ is bilinear in the arguments, $\bm{f}_6(\cdot)$ is an element-wise quadratic function and $\bm{e}''(k) = \text{vec}\left( \bm{E}_R(k) \right)$.

By A\ref{ass:e_noise} and A\ref{ass:z_indep}, it holds that
\begin{align*}
	\bar{E}\left\{\bm{Z}_i(k) \bm{e}(k)\right\} = \bar{E}\left\{\bm{Z}_i(k)\right\} \underbrace{\bar{E}\left\{\bm{e}(k)\right\}}_{=0} = 0,
\end{align*}
and by A\ref{ass:open_loop} it is the case that $\bm{\Omega}_1(k)$,  $\bm{\Omega}_2(k)$, and  $\bm{\Omega}_3(k)$ are independent of $\bm{e}(k)$, $\bm{e}'(k)$ and $\bm{e}''(k)$. Thereby, since each element of $\bm{f}_1(\cdot,\cdot)$,  $\bm{f}_3(\cdot,\cdot)$ and $\bm{f}_5(\cdot,\cdot)$ is linear in the second argument, it can under A\ref{ass:e_noise}, A\ref{ass:open_loop}  and A\ref{ass:z_indep} be concluded that
\begin{align*}
	&\bar{E}\left\{\bm{Z}_i(k) \begin{bmatrix} \bm{f}_1\left(\bm{\Omega}_1(k-1), \bm{e}(k-1)\right) \\ \bm{f}_3\left(\bm{\Omega}_2(k-1), \bm{e}'(k-1)\right)	\\ \bm{f}_5\left(\bm{\Omega}_3(k-1), \bm{e}''(k-1)\right) \end{bmatrix}^T \right\} = 0.
\end{align*}
Moreover, by A\ref{ass:w_noise}, A\ref{ass:z_indep} and A\ref{ass:Ez=0}, it holds that
\begin{align*}
	\bar{E}\left\{\bm{Z}_i(k) \bm{w}(k-1)\right\} = \underbrace{\bar{E}\left\{\bm{Z}_i(k)\right\}}_{=0} \bar{E}\left\{\bm{w}(k-1)\right\}  =  0, 
\end{align*}
and similarly, by A\ref{ass:e_noise}, A\ref{ass:z_indep} and A\ref{ass:Ez=0}, that
\begin{align*}	
	&\bar{E}\left\{\bm{Z}_i(k)  \begin{bmatrix}  \bm{f}_2\left(\bm{e}(k-1)\right) \\ \bm{f}_4\left(\bm{e}'(k-1)\right) \\ \bm{f}_6\left(\bm{e}''(k-1)\right) \end{bmatrix}^T\right\} \\ &= \underbrace{\bar{E}\left\{\bm{Z}_i(k)\right\}}_{=0}  \bar{E}\left\{\begin{bmatrix}  \bm{f}_2\left(\bm{e}(k-1)\right) \\ \bm{f}_4\left(\bm{e}'(k-1)\right) \\ \bm{f}_6\left(\bm{e}''(k-1)\right) \end{bmatrix}^T\right\} = 0.
\end{align*}
Consequently, by A\ref{ass:v_noise}, it is the case that
\begin{align*}
	&\bar{E}\left\{\bm{Z}_i(k) \left( \bm{y}(k) - \hat{\bm{y}}_{2,i}(k, \bm{\theta},  \bm{\rho}, \bm{\lambda}) \right) \right\} \\
	& =  \bar{E}\left\{\bm{Z}_i(k) \begin{bmatrix} \bm{\Phi}\left(\bm{\Omega}_1(k-1)\right) \\ \bm{\Phi}_{\rho, i}\left(\bm{\Omega}_2(k-1)\right)  \\ \bm{\Phi}_{\lambda, i}\left(\bm{\Omega}_3(k-1)\right)  \end{bmatrix}^T \! \begin{bmatrix} \bm{\theta}_0 -  \bm{\theta} \\ \bm{\Omega}_{\rho}(k-1) -  \bm{\rho} \\  \bm{\Omega}_{\lambda}(k-1) -  \bm{\lambda}  \end{bmatrix} \right\}.
\end{align*}
When all experiments, $i  = 1, \hdots, N_E$, are considered, it can by A\ref{ass:informative_data} be concluded that
\begin{align}
	\label{eq:rho_star}\bar{E}\left\{\bm{Z}_i(k) \bm{\Phi}^T_{\rho, i}\left(\bm{\Omega}_2(k-1)\right) \left(\bm{\Omega}_{\rho}(k-1) -  \bm{\rho}\right)  \right\}  = 0, \\
	\label{eq:lambda_star}\bar{E}\left\{\bm{Z}_i(k) \bm{\Phi}^T_{\lambda, i}\left(\bm{\Omega}_3(k-1)\right) \left(\bm{\Omega}_{\lambda}(k-1) -  \bm{\lambda} \right) \right\}  = 0,
\end{align}
have got unique solutions $\bm{\rho}^*$ and $\bm{\lambda}^*$. Consequently, 
\begin{align*}
	\bar{E}\left\{\bm{Z}_i(k) \left( \bm{y}(k) - \hat{\bm{y}}_{2,i}(k, \bm{\theta}_0,  \bm{\rho}^*, \bm{\lambda}^*) \right) \right\} = 0,
\end{align*} 
holds, which means that condition \eqref{eq:IV_cond2} is fulfilled and that the IV method is a consistent estimator of $\bm{\theta}_0$. 
\end{IEEEproof}

\begin{remark}
It is important that the input is informative such that the parameters can be determined uniquely. For example, if both the regressors $x_{\ell} \abs{x_m}$ and $x_m \abs{x_{\ell}}$ are present in $\bm{\Phi}(\cdot)$, both experiments where $x_{\ell}$ and $x_m$ are of the same sign and experiments where they are of opposite sign are needed.
\end{remark}
\begin{remark}
If $\bm{R}(k) = \bar{\bm{R}}$ does not vary with time, A\ref{ass:informative_data} is in general not fulfilled. This means that $\bm{R}(k)$ needs to be excited for the data to be sufficiently informative. 
\end{remark}

Now, supplement A\ref{ass:v_noise} with the following assumption.
\begin{assum} \label{ass:v_noise2}
	The process disturbance $\textbf{v}(k)$ is white, independent of $\textbf{Z}_i(k)$ for $i  = 1, \hdots, N_E$, and independent of $\textbf{w}(\ell)$ for $k \geq \ell$.
\end{assum}
In this case, the following corollary can be proven.
\begin{corollary} \label{corollary1}
	Provided that A\ref{ass:second_order_modulus_system}-A\ref{ass:informative_data} and A\ref{ass:v_noise2} are fulfilled, the IV method defined by \eqref{IV_problem} is a consistent estimator of $\bm{\theta}_0$, $\bar{E}\left\{\bm{\Omega}_{\rho}(k)\right\}$ and $\bar{E}\left\{\bm{\Omega}_{\lambda}(k)\right\}$.
\end{corollary}

\begin{IEEEproof}
By A\ref{ass:open_loop} and A\ref{ass:v_noise2} it is possible to conclude that $\bm{\Omega}_{\rho}(k)$ and $\bm{\Omega}_{\lambda}(k)$ are independent of $\bm{\Omega}_2(k)$, $\bm{\Omega}_3(k)$ and $\textbf{Z}_i(k)$ for $i  = 1, \hdots, N_E$. This means that \eqref{eq:rho_star} and \eqref{eq:lambda_star} can be cast~as
\begin{align*}
	\bar{E}\left\{\bm{Z}_i(k) \bm{\Phi}^T_{\rho, i}\left(\bm{\Omega}_2(k-1)\right)\right\} \bar{E}\left\{ \bm{\Omega}_{\rho}(k-1) -  \bm{\rho}  \right\}  = 0, \\
	\bar{E}\left\{\bm{Z}_i(k) \bm{\Phi}^T_{\lambda, i}\left(\bm{\Omega}_3(k-1)\right)\right\} \bar{E}\left\{\bm{\Omega}_{\lambda}(k-1) -  \bm{\lambda}  \right\}  = 0,
\end{align*}
which have unique solutions $\bm{\rho} = \bar{E}\left\{\bm{\Omega}_{\rho}(k)\right\}$ and $\bm{\lambda} = \bar{E}\left\{\bm{\Omega}_{\lambda}(k)\right\}$. Together with the proof of Theorem~\ref{theorem1}, this shows that the IV method is a consistent estimator of $\bm{\theta}_0$, $\bar{E}\left\{\bm{\Omega}_{\rho}(k)\right\}$ and $\bar{E}\left\{\bm{\Omega}_{\lambda}(k)\right\}$. 
\end{IEEEproof}

\begin{remark}
	For performing undisturbed simulations with system \eqref{eq:system} it is sufficient to know $\bm{\theta}_0$. However, the nuisance parameters, $\bm{\rho}$ and $\bm{\lambda}$, contain information about the first and second-order moments of $\bm{v}(k)$, which can be useful as well, for example for disturbance attenuation by feedforward control.
\end{remark}

\section{Multiple non-additive disturbances}
An interesting extension to the previously studied model class is second-order modulus systems with two non-additive disturbances
\begin{subequations} \label{eq:extended_system}
	\begin{align}
		\nonumber \bm{x}(k+1) &= \bm{f}_{\text{som}, 1}\left( \begin{bmatrix}  \bm{x}(k) + \bm{R}(k)\bm{v}_1(k) \\ \bm{u}(k)  \end{bmatrix}, \bm{\theta}_{0,1} \right) \\ 
		& + \bm{f}_{\text{som}, 2}\left( \begin{bmatrix} \bm{x}(k) + \bm{R}(k)\bm{v}_2(k) \\ \bm{u}(k)  \end{bmatrix}, \bm{\theta}_{0,2} \right) + \bm{w}(k) \\
		\bm{y}(k) &= \bm{x}(k) + \bm{e}(k), \\
		\bm{Y}_R(k) &= \bm{R}(k) + \bm{E}_R(k).
	\end{align}
\end{subequations}
Systems like this appear in practice when vessels are moving in two surrounding media at once, which is the case for marine surface vessels.

\subsection{Identifiability and informativity discussion}
It can immediately be noted that the system formulation, \eqref{eq:extended_system}, is problematic when $\bm{f}_{\text{som}, 1}(\cdot)$ and $\bm{f}_{\text{som}, 2}(\cdot)$ have common terms. Actually, unless additional information is provided, unique identification of parameters associated with common terms is not possible. A set of predictors such as \eqref{eq:augmented_predictor} and the estimation framework suggested in the previous section can still be applied, but in this case $\bm{\rho}$ and $\bm{\lambda}$ will asymptotically depend on mixed moments of $\bm{v}_1(k)$ and $\bm{v}_2(k)$, whereas $\bm{\theta}$ in general will converge to an aggregation of $\bm{\theta}_{0,1}$ and $\bm{\theta}_{0,2}$. In order to uniquely identify $\bm{\theta}_{0,1}$ and $\bm{\theta}_{0,2}$, an auxiliary disturbance measurement can be considered.
\begin{assum} \label{ass:disturbance_measurement}
	There is an auxiliary measurement, $\bm{y}_{\text{aux}}(k) = \bm{R}(k)\bm{v}_2(k) + \bm{e}_{\text{aux}}(k)$, available. The associated measurement noise, $\bm{e}_{\text{aux}}(k)$, is a stationary signal with zero mean and well-defined moments of any order.
\end{assum}
Then, the set of predictors
\begin{align} \label{eq:dist_meas_predictor}
	\nonumber	\hat{\bm{y}}_{3,i}(k, \bm{\theta}_1, \bm{\theta}_2, \bm{\rho}, \bm{\lambda}) &= 	\hat{\bm{y}}_{2,i}(k, \bm{\theta}_1, \bm{\rho}, \bm{\lambda}) \\
	& \hspace{-2cm}+ \bm{\Phi}_2^T\left( \begin{bmatrix}  \bm{y}(k-1) + \bm{y}_{\text{aux}}(k-1) \\ \bm{u}(k-1)  \end{bmatrix} \right)  \bm{\theta}_2,
\end{align}
can be used. Here $\hat{\bm{y}}_{2,i}(k, \bm{\theta}_1, \bm{\rho}, \bm{\lambda})$ comes from $\bm{f}_{\text{som}, 1}(\cdot)$ as before and $\bm{\Phi}_2(\cdot)$ is the matrix associated with $\bm{f}_{\text{som}, 2}(\cdot)$ by Definition~\ref{def:modulus}. Notably, there can still be issues with common terms. The case when $\bm{f}_{\text{som}, 1}(\cdot)$ and $\bm{f}_{\text{som}, 2}(\cdot)$ have common terms which only depend on $\bm{u}(k)$ is not that interesting and can typically be dealt with by reformulating the problem. However, common terms that depend on the state can be present in \eqref{eq:extended_system}, in which case there will be issues if $\bm{R}(k)\bm{v}_1(k) = \bm{R}(k)\bm{v}_2(k)$ for all $k$. This happens if $\bm{v}_1(k) = \bm{v}_2(k)$ but can happen under milder conditions if $\bm{R}(k)$ does not have full column rank. Moreover, if the measured non-additive disturbance does not vary with time, $\bm{v}_2(k) = \bar{\bm{v}}_2$, the auxiliary measurement, $\bm{y}_{\text{aux}}(k)$, does not provide sufficiently much new information beyond $\bm{y}(k)$ and $\bm{Y}_R(k)$, which has as consequence that some common terms cannot be distinguished from each other. Based on these observations, the following assumptions are made regarding identifiability and data informativity.

\begin{assum} \label{ass:input_terms_identifiability}
	The functions $\bm{f}_{\text{som}, 1}(\cdot)$ and $\bm{f}_{\text{som}, 2}(\cdot)$ do not have common terms that only depend on $\bm{u}(k)$.
\end{assum}

\begin{assum} \label{ass:disturbance_informativity}
	The non-additive disturbances are such that $\bm{R}(k)\bm{v}_1(k) \neq \bm{R}(k)\bm{v}_2(k)$ and $\bm{v}_2(k) \neq \bm{v}_2(\ell)$ for almost all $k$ and $\ell$.
\end{assum}

\begin{assum} \label{ass:extra_instrument_matrix}
	For each experiment $i = 1, \hdots, N_E$, there exists an instrument matrix $\bm{Z}_i(k)$ which, in addition to fulfilling A\ref{ass:z_indep}-A\ref{ass:informative_data}, is independent of $\bm{e}_{\text{aux}}(k)$ and for which $\bar{E}\left\{ \bm{Z}_i(k) \bm{\Phi}_2^T(\cdot) \right\}$ has full rank.
\end{assum}

\subsection{Main result}
Define the IV estimator as the least-squares solution to
\begin{align} \label{IV_problem2}
	\begin{cases} \frac{1}{N_D}\sum\limits_{k=1}^{N_D} \bm{Z}_1(k) \left(\bm{y}(k)-\hat{\bm{y}}_{3,1}(k, \bm{\theta}_1, \bm{\theta}_2, \bm{\rho}, \bm{\lambda})\right) = 0, \\ \qquad \qquad \qquad \qquad \qquad\vdots \\  \frac{1}{N_D} \hspace{-0.4cm} \sum\limits_{k=(N_E-1)N_D+1}^{N_E N_D} \hspace{-0.6cm} \bm{Z}_{N_E}(k) \left(\bm{y}(k)\!-\!\hat{\bm{y}}_{3, N_E}(k, \bm{\theta}_1, \bm{\theta}_2, \bm{\rho}, \bm{\lambda})\right) \! = \! 0,  \end{cases}
\end{align}
and supplement A\ref{ass:excitation_offset} with a final assumption.
\begin{assum} \label{ass:excitation_offset2}
	The input in each experiment excites the system such that each of its states continuously has an amplitude that fulfills
	\begin{align*}
		\abs{x_i(k) + \sum_{j=1}^{n_v} \bm{R}_{i,j}(k) v_{2,j}(k)} > \abs{e_i(k) + e_{\emph{aux},i}(k)},
	\end{align*}
	for $k = 1, \ \hdots \ N_D$, $i = 1, \ \hdots \ n_x$.
\end{assum}

\begin{theorem} \label{theorem2}
		Provided that data is generated based on \eqref{eq:extended_system}, that A\ref{ass:second_order_modulus_system}-A\ref{ass:informative_data} hold (for $\bm{v}_1(k)$ in place of $\bm{v}(k)$) and that A\ref{ass:disturbance_measurement}-A\ref{ass:excitation_offset2} hold,  the IV method defined by \eqref{IV_problem2} is a consistent estimator of $\bm{\theta}_{0,1}$ and $\bm{\theta}_{0,2}$.
\end{theorem}

\begin{IEEEproof}
An IV estimator is consistent if conditions \eqref{eq:IV_cond1} and \eqref{eq:IV_cond2} hold. Condition \eqref{eq:IV_cond1} will hold if there exists a matrix $\bm{Z}_{i}(k)$ for $i = 1, \hdots, N_E$, such that
\begin{align*}
	\bm{H} &= \begin{bmatrix} \bm{H}_1 & \hdots & \bm{H}_{N_E} \end{bmatrix}^T,
\end{align*}
with block elements
\begin{align*}
	\bm{H}_i &= \Bar{E} \left\{ \bm{Z}_{i}(k) \begin{bmatrix} \bm{\Phi}^T( \cdot ) & \bm{\Phi}^T_{\rho, i}(\cdot) & \bm{\Phi}^T_{\lambda, i}(\cdot) & \bm{\Phi}_2^T( \cdot ) &  \end{bmatrix} \right\},
\end{align*}
has full rank. By A\ref{ass:extra_instrument_matrix} it is the case that $\bar{E}\left\{ \bm{Z}_i(k) \bm{\Phi}_2^T(\cdot) \right\}$ has full rank. Moreover, under A\ref{ass:input_terms_identifiability} and A\ref{ass:disturbance_informativity}, the columns of $\bm{\Phi}_2^T( \cdot )$ are linearly independent of the columns of $\bm{\Phi}^T( \cdot )$, $\bm{\Phi}^T_{\rho, i}(\cdot)$ and  $\bm{\Phi}^T_{\lambda, i}(\cdot)$. These observations combined with A\ref{ass:informative_data} makes it possible to conclude that $\bm{H}$ will have full rank, which in turn means that  \eqref{eq:IV_cond1} is fulfilled.

Now it will be shown that \eqref{eq:IV_cond2} holds for $i = 1, \hdots, N_E$. From \eqref{eq:extended_system}, it follows that
\begin{align*}
	\bm{y}(k) &= \underbrace{\bm{f}_{\text{som}, 1}(\cdot) + \bm{w}(k-1) + \bm{e}(k)}_{= \bm{\xi}_1(k)} + \underbrace{\bm{f}_{\text{som}, 2}(\cdot)}_{= \bm{\xi}_2(k)},
\end{align*}
and the fact that
\begin{align*}
	\bar{E}\left\{ \bm{Z}_i(k) \left(\bm{\xi}_1(k) - \hat{\bm{y}}_{2, i}(k, \bm{\theta}_1, \bm{\rho}, \bm{\lambda}) \right) \right\} &= 0,
\end{align*}
is fulfilled for $\bm{\theta} = \bm{\theta}_{0,1}$, $\bm{\rho} = \bm{\rho}^*$ and $\bm{\lambda} = \bm{\lambda}^*$, was shown in the proof of Theorem~\ref{theorem1}. Thereby, in view of \eqref{eq:dist_meas_predictor} it remains to show that
\begin{align*}
	\bar{E}\left\{ \bm{Z}_i(k) \left(\bm{\xi}_2(k) - \bm{\Phi}_2^T(\cdot) \bm{\theta}_{0,2} \right) \right\} &= 0,
\end{align*}
or more specifically that
\begin{align} \label{eq:proof2_condition}
	&\bar{E}\!\left\{\!\bm{Z}_i(k) \left(\bm{\Phi}_2^T(\bm{\Omega}_4(k-1)) \!-\! \bm{\Phi}_2^T(\bm{\Omega}'_4(k-1)) \right) \!\right\}\bm{\theta}_{0,2} = 0,
\end{align}
holds, where
\begin{align*}
	\bm{\Omega}_4(k) \! = \!\begin{bmatrix} \bm{x}(k) + \bm{R}(k)\bm{v}_2(k) \\ \bm{u}(k)  \end{bmatrix}\!, \ \bm{\Omega}'_4(k) \! = \!\begin{bmatrix}  \bm{y}(k) + \bm{y}_{\text{aux}}(k) \\ \bm{u}(k)  \end{bmatrix}
\end{align*}
It can be noted that
\begin{align*}
	\bm{\Omega}'_4(k) = \bm{\Omega}_4(k) + \begin{bmatrix}  \bm{e}(k) + \bm{e}_{\text{aux}}(k) \\ 0  \end{bmatrix},
\end{align*}
and that
\begin{align*}
	\bar{E}\left\{\bm{Z}_i(k)\left(\bm{e}(k-1) + \bm{e}_{\text{aux}}(k-1) \right)\right\} = 0,
\end{align*}
holds under A\ref{ass:e_noise}, A\ref{ass:z_indep}, A\ref{ass:disturbance_measurement} and A\ref{ass:extra_instrument_matrix}. Moreover, the sign of all modulus expressions are known by A\ref{ass:excitation_offset2}, which means that $\bm{\Phi}_2(\cdot)$ temporarily can be treated as a second-order function. When the squares are developed, it is possible to express $\bm{\Phi}_2\left(\bm{\Omega}'_4(k)\right)$ as
\begin{align*}
	\bm{\Phi}_2\left(\bm{\Omega}_4(k)\right) + \bm{f}_7\left(\bm{\Omega}_4(k), \bm{e}'''(k) \right) + \bm{f}_8\left(\bm{e}'''(k)\right),
\end{align*}
where each element of $\bm{f}_7(\cdot,\cdot)$ is bilinear in the arguments, $\bm{f}_8(\cdot)$ is an element-wise quadratic function and $\bm{e}'''(k) = \bm{e}(k) + \bm{e}_{\text{aux}}(k)$. Thereby, reasoning analogously to the proof of Theorem~\ref{theorem1}, it is possible to conclude~that
\begin{align*}
	\bar{E}\left\{\bm{Z}_i(k) \bm{f}^T_7\left(\bm{\Omega}_4(k), \bm{e}'''(k) \right) \right\} = 0,
\end{align*}
holds by A\ref{ass:e_noise}, A\ref{ass:open_loop}, A\ref{ass:z_indep}, A\ref{ass:disturbance_measurement}, A\ref{ass:extra_instrument_matrix} and that
\begin{align*}	
	\bar{E}\left\{\bm{Z}_i(k) \bm{f}^T_8\left(\bm{e}'''(k)\right) \right\} = 0.
\end{align*}
holds by A\ref{ass:e_noise}, A\ref{ass:z_indep}, A\ref{ass:disturbance_measurement} and A\ref{ass:extra_instrument_matrix}. Consequently, \eqref{eq:proof2_condition} holds true, which in turn means that \eqref{eq:IV_cond2} is fulfilled. This shows that the IV method is a consistent estimator of $\bm{\theta}_{0,1}$ and $\bm{\theta}_{0,2}$.
\end{IEEEproof}

\begin{remark}
	Whether it is feasible to measure a system disturbance or not is highly application dependent. For example, ships are often equipped with sensors for measuring wind speed and direction but rarely with sensors for measuring speed and direction of ocean currents.
\end{remark}

\begin{remark}
	A\ref{ass:excitation_offset2} can be more or less restricting than A\ref{ass:excitation_offset}. In particular for disturbances with large magnitude, A\ref{ass:excitation_offset2} is easier to fulfill. This is further illustrated in the subsequent simulation example.
\end{remark}

\begin{remark}
	The measured non-additive process disturbance, $\bm{v}_2(k)$, does not need to be a stationary signal and can include a deterministic time-dependent component. 
\end{remark}

\section{Simulation example}
In order to illustrate the potential of the proposed estimators, simulation experiments were performed. These were carried out using a model of a surface vessel. For simplified notation, the time dependence of continuous-time signals will not be written out explicitly.

\subsection{Ship modelling}
Physical modelling of marine vessels is a complicated matter and the ambition with this work was not to develop new theory in that regard. Therefore, the modeling framework in \cite{fossen2011handbook} was adopted. There, the equations of motion are, as in \eqref{eq:newton2}, derived based on superposition of forces in accordance with Newton's laws and by convention expressed on matrix form
\begin{subequations} \label{eq:equations_of_motion}
\begin{align}
	\label{eq:equations_of_motion_A} 
	\dot{\bm{\eta}} = \bm{J}(\bm{\eta}) \bm{\nu},  \\
	\label{eq:equations_of_motion_B}
	\bm{M} \dot{\bm{\nu}} + \bm{C}(\bm{\nu}_r) \bm{\nu}_r + \bm{D}(\bm{\nu}_r) \bm{\nu}_r + \bm{F}(\bm{\nu}_q) \bm{\nu}_q = \bm{\tau}. 
\end{align}
\end{subequations}
Here the first state vector, $\bm{\eta}$, constitutes global position and attitude in the form of Euler angles between an inertial frame and the body-fixed frame. The second state vector, $\bm{\nu}$, includes translational velocities expressed in the body-fixed frame and angular velocities between the two frames. The subscripts signify relative velocities such that $\bm{\nu}_r = \bm{\nu} -\bm{\nu}_c$ and $\bm{\nu}_q = \bm{\nu} - \bm{\nu}_w$, where $\bm{\nu}_c$ is the velocity of an ocean current and $\bm{\nu}_w$ is the wind velocity. Moreover, $\bm{J}(\bm{\eta})$ is an attitude dependent rotation matrix, $\bm{M}$ is a matrix including mass and inertia elements, $\bm{C}(\cdot)$ captures Coriolis and centripetal effects and $\bm{D}(\cdot)$ describes energy losses due to hydrodynamic damping. The structures of all these matrix functions were taken from (7.10), (7.13), (7.16), (7.19) and (7.24) in \cite{fossen2011handbook}, with the additional assumption of a short distance between the ship's center of gravity and the origin of the body-fixed coordinate system. These choices give what is called a maneuvering model, $i.e.$, a model where only motion in the horizontal plane is considered and where the dynamics associated with motion in heave, pitch and roll are neglected. This approximation is usually justifiable for large vessels with flat keels and give
\begin{align*}
	\bm{\eta} = \begin{bmatrix} x & y & \psi	\end{bmatrix}^T, \qquad \bm{\nu} = \begin{bmatrix} u & v & r	\end{bmatrix}^T,
\end{align*}
where $u$, $v$, and $r$ constitute surge, sway and yaw rate, respectively, whereas $x$ and $y$ are positional coordinates in the horizontal plane and $\psi$ is the heading angle.

Marine surface vessels move in air and water at the same time, but it is common to neglect the fact that the aerodynamic forces depend on the velocity of the ship. This is a reasonable approximation in some cases because aerodynamic forces and moments are significantly much smaller than their hydrodynamic counterparts. However, in general, wind effects will be nonlinear and enter both additively and multiplicatively in the equations of motion. In this work, $\bm{F}(\cdot)$ was used to describe energy losses due to aerodynamic drag, which includes effects of the wind as well as regular air resistance. In theory, there will be both added-mass and Coriolis effects connected to the moved air as well, but these effects are supposedly small and a damping matrix was assumed to be sufficient for capturing all relevant aerodynamic effects. The aerodynamic damping forces, $\bm{F}(\bm{\nu}_q) \bm{\nu}_q$, were assumed to be given by second-order modulus expressions similar to (7.24). Lastly, the system input $\bm{\tau}$ is a collection of forces and moments caused by the ship's actuators. These input forces were treated as known signals and no actuator dynamics were considered to keep the simulation example simple and transparent.

If $\bm{M}$ is non-singular, \eqref{eq:equations_of_motion_B} can be cast on state-space form
\begin{align*}
	 \dot{\bm{\nu}} &= \bm{M}^{-1}\left( -\bm{C}(\bm{\nu}_r) \bm{\nu}_r - \bm{D}(\bm{\nu}_r) \bm{\nu}_r - \bm{F}(\bm{\nu}_q) \bm{\nu}_q + \bm{\tau} \right), 
\end{align*}
which, with this choice of matrix functions, has second-order modulus nonlinearities. Since vessel models are often based on physical principles, they are usually first formulated in continuous time like this. However, a continuous-time second-order modulus model can be cast as a discrete-time model with preserved nonlinear structure, using for example Euler's explicit method. If the sampling frequency is sufficiently much faster than the frequency of the signal variations, the accuracy of this approximation will be good. The dynamics of large surface vessels are usually slow in comparison to the sampling of the measured signals, which makes an Euler approximation sufficiently accurate. 

\begin{table}[hb]
\begin{center}
\caption{System premises for simulation.}
\label{table:system_parameters}
\begin{tabular}{||c|c|c||}
			\hline
			Parameter &  True system & Nominal model                          	\\  \hline
			$\mathcal{X}_u$              	& $-0.05$    	& $-0.2$          	\\  \hline
			$\mathcal{X}_{vr}$              & $1$   		& $0.8$           	\\  \hline
			$\mathcal{X}_{|u|u}$            & $-0.05$    	& $0$           	\\  \hline
			$\mathcal{W}_{|u|u}$            & $-0.0005$ 	& $0$        		\\  \hline
			$\mathcal{X}_{\tau}$             & $0.02$      	& $0.01$            \\  \hline
			$\mathcal{Y}_v$              	& $-0.2$      	& $-0.3$            \\  \hline
			$\mathcal{Y}_{ur}$              & $-0.65$  		& $-0.8$           	\\  \hline
			$\mathcal{Y}_{|v|v}$           	& $-0.2$   		& $0$           	\\  \hline
			$\mathcal{Y}_{|v|r}$           	& $-0.1$   		& $0$            	\\  \hline
			$\mathcal{W}_{|v|v}$           	& $-0.0015$  	& $0$        		\\  \hline
			$\mathcal{Y}_{\tau}$           	& $0.02$     	&  $0.01$         	\\  \hline
			$\mathcal{N}_r$           		& $-0.1$       	& $-0.15$           \\  \hline
			$\mathcal{N}_{uv}$           	& $-0.0015$ 	& $0$ 				\\  \hline
			$\mathcal{N}_{|v|v}$           	& $-0.001$  	& $0$            	\\  \hline
			$\mathcal{N}_{|v|r}$           	& $-0.04$   	& $0$           	\\  \hline
			$\mathcal{W}_{uv}$           	& $-0.00003$   	& $0$           	\\  \hline
			$\mathcal{N}_{\tau}$           	& $0.0003$   	& $0.00015$         \\ 
			\hline      
\end{tabular}		
\end{center}
\end{table}

After discretization, with unit sampling time for brevity, the model structure
\begin{align*} 
	\begin{bmatrix} u(k) \\ v(k) \\ r(k) \end{bmatrix} =  \begin{bmatrix} u(k-1) \\ v(k-1) \\ r(k-1) \end{bmatrix} + \begin{bmatrix} \bm{\varphi}^T_{u}(k) & 0 & 0 \\ 0 & \bm{\varphi}^T_{v}(k) & 0 \\ 0 & 0 & \bm{\varphi}^T_{r}(k) \end{bmatrix} \bm{\theta_0}.
\end{align*}
is obtained, where
\begin{align*}
	\bm{\varphi}_{u}(k) &=
	\left[\begin{matrix}
		u_r(k\!-\!1) &   v_r(k-1)r(k\!-\!1) & u_r(k\!-\!1) \abs{u_r(k\!-\!1)} 
	\end{matrix}\right.\\
	&
	\left.\begin{matrix}
		 u_q(k\!-\!1) \abs{u_q(k\!-\!1)} & \tau_{1}(k\!-\!1)  
	\end{matrix}\right]^T,
\end{align*}
\begin{align*}
	\bm{\varphi}_{v}(k) &=
	\left[\begin{matrix}
		v_r(k\!-\!1) &  u_r(k\!-\!1)r(k\!-\!1) & v_r(k\!-\!1) \abs{v_r(k\!-\!1)} 
	\end{matrix}\right.\\
	&
	\left.\begin{matrix}
		 r(k\!-\!1) \abs{v_r(k\!-\!1)} & v_q(k\!-\!1) \abs{v_q(k\!-\!1)} & \tau_{2}(k\!-\!1)
	\end{matrix}\right]^T,
\end{align*}
\begin{align*}
	\bm{\varphi}_{r}(k) &=
	\left[\begin{matrix}
		r(k\!-\!1) & u_r(k\!-\!1)v_r(k\!-\!1) & v_r(k\!-\!1) \abs{v_r(k\!-\!1)} 
	\end{matrix}\right.\\
	&
	\left.\begin{matrix}
		 r(k\!-\!1) \abs{v_r(k\!-\!1)} & u_q(k\!-\!1)v_q(k\!-\!1) & \tau_{3}(k\!-\!1)
	\end{matrix}\right]^T,
\end{align*}
and
\begin{align*}
	 \bm{\theta}_0 &=
	\left[\begin{matrix}
		\mathcal{X}_u \! & \mathcal{X}_{vr} \! & \mathcal{X}_{|u|u} \! & \mathcal{W}_{|u|u} \! &  \mathcal{X}_{\tau} \! &  \mathcal{Y}_v \! & \mathcal{Y}_{ur} \! & \mathcal{Y}_{|v|v} \! & \mathcal{Y}_{|v|r} \!
	\end{matrix}\right.\\
	& \hspace{0.1cm}
	\left.\begin{matrix}
		 & \mathcal{W}_{|v|v} & \mathcal{Y}_{\tau} & \mathcal{N}_r & \mathcal{N}_{uv} & \mathcal{N}_{|v|v} & \mathcal{N}_{|v|r} & \mathcal{W}_{uv} & \mathcal{N}_{\tau}
	\end{matrix}\right]^T.
\end{align*}
Marine models are often overparameterized and unique identification of all physical effects is not necessarily possible in practice. The curly notation for the elements of $\bm{\theta}_0$ above, indicates that the original model parameters in \cite{fossen2011handbook} were redefined to get identifiability. The parameter values used to generate data in the simulation experiments can be seen in the second column of Table~\ref{table:system_parameters}. These values were chosen based on earlier works with experimental data from a full-scale marine vessel and an interested reader will find more details in \cite{ljungberg2020lic}.

The wind and ocean currents are modelled as stationary stochastic processes in the inertial frame. This means that, when expressed in the body-fixed frame, they depend on the attitude of the ship
\begin{align*}
	\bm{\nu}_c(k) &= \bm{J}^{-1}\left(\bm{\eta}(k)\right) \underbrace{\begin{bmatrix}\nu_{c, NS}(k) & \nu_{c, EW}(k) & 0 \end{bmatrix}^T}_{=\bm{\nu}_c'(k)}, \\ \bm{\nu}_w(k) &= \bm{J}^{-1}\left(\bm{\eta}(k)\right) \underbrace{\begin{bmatrix}\nu_{w, NS}(k) & \nu_{w, EW}(k) & 0 \end{bmatrix}^T}_{=\bm{\nu}_w'(k)}.
\end{align*}
Here the subscripts NS and EW indicate north/south and east/west components, respectively. For horizontal motion of a vessel, the kinematic equations, \eqref{eq:equations_of_motion_A}, reduce to one principal rotation about the vertical axis. In this case, the rotation matrix solely depends on the yaw angle, $\bm{J}(\bm{\eta}) = \bm{J}(\psi)$. In addition to the velocity states, it was assumed that this angle was measured
\begin{align*}
	\bm{y}_{\nu}(k) &= \bm{\nu}(k) + \bm{e}_{\nu}(k), \\
	y_{\psi}(k) &= \psi(k) + e_{\psi}(k).
\end{align*}
Then, it is for each time instant possible to estimate the rotation matrix, and more importantly its inverse
\begin{align*}
	\hat{\bm{R}}(k) = \bm{J}^{-1}\left(y_{\psi}(k)\right).
\end{align*}
In summary, a system representation similar to the one in \eqref{eq:extended_system} is obtained.

\subsection{Simulation setup}
Four estimators were compared in the simulation experiments. Three of these were IV estimators that differed from each other by being based on different predictors. The first one, $\hat{\bm{\theta}}_N^{IV_1}$, was based on the basic predictor~\eqref{eq:basic_predictor}, the second one, $\hat{\bm{\theta}}_N^{IV_2}$, on the set of augmented predictors defined by~\eqref{eq:augmented_predictor} and the third one, $\hat{\bm{\theta}}_N^{IV_3}$, on the set of predictors in~\eqref{eq:dist_meas_predictor}, where auxiliary wind measurements
\begin{align*}
	\bm{y}_{\text{aux}}(k) = \bm{J}^{-1}\left(\psi(k)\right)  \bm{\nu}_w'(k) + \bm{e}_{\text{aux}}(k),
\end{align*}
were utilized. To make it clear that an estimator based on PEM, using any of the aforementioned predictors, for significant amounts of data is inaccurate, the results of a least-squares (LS) estimator were included too. This LS estimator, $\hat{\bm{\theta}}_N^{LS}$, was based on the set of predictors in~\eqref{eq:dist_meas_predictor} (LS estimators based on the other predictors gave less accurate models).

In preceding sections, the existence of instrument matrices with certain properties have been assumed, but no explanations have been given for how such matrices can be found. A common way of obtaining instruments in practice is by simulation of a nominal model with crude parameter values. In this case, the instrument matrix can be formed as a noise-free version of the regression matrix. The parameters can then be refined by iteratively letting the instruments be simulated from the model parameterized by the latest set of parameters until convergence, as described in \cite{young2008refined}. Here, the nominal model had the same structure as the true system and its parameters are given in the right column of Table~\ref{table:system_parameters}.  All the IV estimators use zero-mean instruments and to obtain this, the average value of each component of the instrument matrix was simply subtracted.

In the simulation experiments, all disturbances were sampled from Gaussian distributions
\begingroup
\allowdisplaybreaks
\begin{align*}
	\bm{e}_{\nu}(k) &\sim \mathcal{N}(0, 2 \cdot 10^{-4} \cdot \mathcal{I}_3), \\
	e_{\psi}(k) &\sim \mathcal{N}(0, 10^{-4}), \\
	\begin{bmatrix}	\nu_{c, NS}(k) & \nu_{c, EW}(k) \end{bmatrix}^T &\sim \mathcal{N}\left(\begin{bmatrix}	0.2 & 0.2 \end{bmatrix}^T, 10^{-3} \cdot \mathcal{I}_2\right),	\\
	\begin{bmatrix}	\nu_{w, NS}(k) & \nu_{w, EW}(k) \end{bmatrix}^T &\sim \mathcal{N}\left(\begin{bmatrix}	\bar{\nu}_w & \bar{\nu}_w \end{bmatrix}^T, 10^{-3} \cdot \mathcal{I}_2\right), \\
	\bm{e}_{\text{aux}}(k) &\sim \mathcal{N}(0, 10^{-3} \cdot \mathcal{I}_3), \\	
\end{align*}
\endgroup
Here $\mathcal{I}_n$ is the $n \times n$ identity matrix. Moreover, the input signal was divided into two parts
\begin{align*}
	\bm{\tau}(k) &= \tilde{\bm{\tau}}(k) + \bar{\bm{\tau}},
\end{align*}
where the time-varying component, $\tilde{\bm{\tau}}(k)$, was smoothed pulses of varying width that excited the system well and the static component, $\bar{\bm{\tau}}$, was chosen such that the ship had a positive surge and sway speed. Applying the static part of the input alone made the ship move in a wide circle. Consequently, when the time-varying part was added on top, the ship made small zigzag-like deviations from this main circle. An experiment design like this makes the data fulfill the requirements of having the surge and sway states well-separated from the origin, simultaneously. Notably, the yaw rate, $r(k)$, does not need to be of any particular sign because it never appears inside a modulus function.

\subsection{Simulation results}
In situations with limited amounts of estimation data, an inconsistent estimator might very well give better-performing models than a consistent one. To evaluate the variance properties of the suggested estimators, simulations experiments were therefore performed where the amount of estimation data was varied. The results of these simulations were assessed using the normalized model-fit metric, see for example \cite{ljung1999system}. This metric is sometimes interpreted as a percentage because the best possible outcome is 100. The output of a model can, however, have an arbitrarily bad fit and values below 0 are therefore possible.

\begin{figure}
	\centering
	\includegraphics[width = 8.4cm]{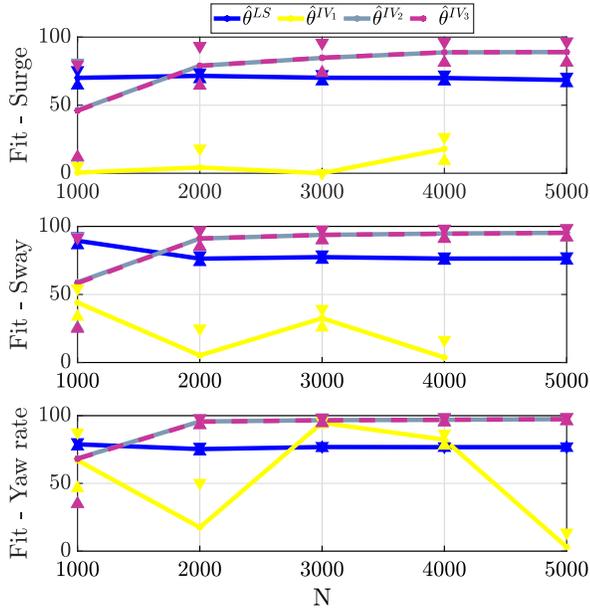}
	\caption{Average fit of models obtained from the four estimators for different values of $N$ when data with low wind speed was used for estimation. The triangles indicate average model fit plus/minus one standard deviation.}
	\label{fig:low_wind_fit}
\end{figure}

\begin{figure}
	\centering
	\includegraphics[width = 8.4cm]{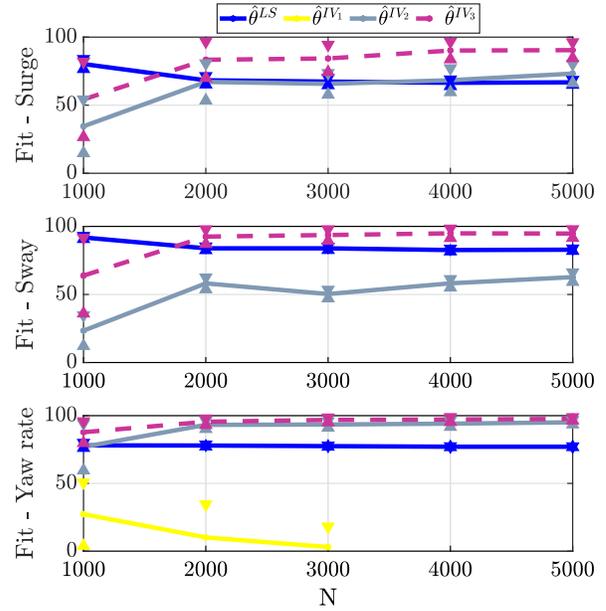}
	\caption{Average fit of models from the four estimators for different values of $N$ when data with high wind speed was used for estimation. Triangles indicate average model fit plus/minus one standard deviation.}
	\label{fig:high_wind_fit}
\end{figure}

The simulations were divided into two cases, one with moderately low wind speed, $\bar{\nu}_w = 1 \ \text{m/s}$, and one with high wind speed, $\bar{\nu}_w = 10 \ \text{m/s}$. The model fit was calculated by comparing the simulated response with an undisturbed set of validation data. In the validation dataset, a more conventional input signal was used, which made the ship move forward in a zigzag manner. The results are given in  Figures~\ref{fig:low_wind_fit} and \ref{fig:high_wind_fit} for the two cases, respectively. The figures were obtained by averaging the results of 100 Monte Carlo iterations for different values of $N$ between $1000$ and $5000$. The average values of fit plus/minus one standard deviation are marked with triangles. In Figure~\ref{fig:low_wind_fit} it can be seen that for low amounts of data, the models obtained from $\hat{\bm{\theta}}_N^{LS}$ are performing better than models from the IV estimators. However, after a certain breakpoint, the accuracy of models obtained with the estimators $\hat{\bm{\theta}}_N^{IV_2}$ and $\hat{\bm{\theta}}_N^{IV_3}$ seem to overtake the accuracy of those obtained with the LS estimator. Moreover, in the case of low wind speed, the estimators $\hat{\bm{\theta}}_N^{IV_2}$ and $\hat{\bm{\theta}}_N^{IV_3}$ are performing equally well, but in the case of high wind speed, the short-comings of $\hat{\bm{\theta}}_N^{IV_2}$ are clear, especially in surge. This is because the wind speed is higher than the speed of the ship, which means that A\ref{ass:excitation_offset} is violated in terms of the wind disturbance. Notably, A\ref{ass:excitation_offset2} is still fulfilled in this case, which is the reason that the estimator $\hat{\bm{\theta}}_N^{IV_3}$ is performing well. The estimator, $\hat{\bm{\theta}}_N^{IV_1}$, generates models with poor model fit and to make the other results clear, the figures are retained to only show values of model fit between 0 and 100. As a consequence, the results of that estimator are sometimes not shown.

\section{Conclusions}
Consistent parameter estimators for second-order modulus models have been developed. These results are relevant in robotic applications, where quadratic nonlinearities are needed for reproducing essential aspects of system behaviors. There is a close connection between this work and the design of disturbance observers. Potential future work is to see how accurately the first and second-order moments of the environmental disturbances can be estimated in the case where a model of the undisturbed system is already known.

\bibliographystyle{plain}        
\bibliography{bib}           	 

\end{document}